\documentclass[iop, apjl]{emulateapj}
\usepackage{threeparttable}
\usepackage{multirow}
\usepackage{times}
\usepackage{enumitem}
\usepackage{url} 
\usepackage{color}
\usepackage{amsmath,bm}

\shorttitle{}
\shortauthors{Y. Huang et al.}

\begin{document}

\title{On the kinematic signature of the Galactic warp as revealed by the LAMOST-TGAS data}

\author{Y. Huang\altaffilmark{1,2,7}}
\author{ R. Sch{\"o}nrich\altaffilmark{3}}
\author{ X.-W. Liu\altaffilmark{1}}
\author{ B.-Q. Chen\altaffilmark{1}}
\author{ H.-W. Zhang\altaffilmark{2,4}}
\author{ H.-B. Yuan\altaffilmark{5}}
\author{ M.-S. Xiang\altaffilmark{6,7}}
\author{ C. Wang\altaffilmark{2}}
\author{ Z.-J. Tian\altaffilmark{2,7}}

\altaffiltext{1}{South-Western Institute for Astronomy Research, Yunnan University, Kunming 650500, People's Republic of China; {\it yanghuang@pku.edu.cn {\rm (YH)}; x.liu@pku.edu.cn {\rm (XWL)}}}
\altaffiltext{2}{Department of Astronomy, Peking University, Beijing 100871, People's Republic of China {\it zhanghw@pku.edu.cn (\rm HWZ)}}
\altaffiltext{3}{Rudolf Peierls Centre for Theoretical Physics, 1 Keble Road, Oxford, OX1 3NP, UK; {\it ralph.schoenrich@physics.ox.ac.uk}}
\altaffiltext{4}{Kavli Institute for Astronomy and Astrophysics, Peking University, Beijing 100871, People's Republic of China}
\altaffiltext{5}{Department of Astronomy, Beijing Normal University, Beijing 100875, People's Republic of China}
\altaffiltext{6}{Key Laboratory of Optical Astronomy, National Astronomical Observatories, Chinese Academy of Sciences, Beijing 100012, People's Republic of China}
\altaffiltext{7}{LAMOST Fellow}

\begin{abstract}
Using a sample of about 123,000  stars with accurate 3D velocity measurements from the LAMOST-TGAS data, we confirm the kinematic signature of the Galactic warp found by Sch{\"o}nrich \& Dehnen recently.
The data reveal a clear trend of increasing mean vertical velocity $\overline{V_{z}}$ as a function of absolute vertical angular momentum $L_{z}$ and azimuthal velocity $V_{\phi}$ for guiding center radius $R_{g}$ between 6.0 and 10.5\,kpc.
The trend is consistent with a {large-scale} Galactic warp.
Similar to Sch{\"o}nrich \& Dehnen, we also find a wave-like pattern of  $\overline{V_{z}}$ versus $L_{z}$ with an amplitude of $\sim 0.9$\,km\,s$^{-1}$ on a scale of $\sim 2.0$\,kpc, which could arise from bending waves or a winding warp.
Finally, we confirm a prominent, localized peak in $\overline{V_z}$ near $L_z \sim 2150$\,kpc\,km\,s$^{-1}$ (corresponding to $R_{g} \sim 9$\,kpc  and $V_{\phi} \sim 255$\,km\,s$^{-1}$). 
The additional line-of-sight velocity information from LAMOST reveals that stars in this feature have a large, { inward} radial velocity of $V_{R} \sim -13.33 \pm 0.59$\,km\,s$^{-1}$ and a small radial velocity dispersion $\sigma_{R} \sim 25.27 \pm 0.89$\,km\,s$^{-1}$, suggesting that a stellar stream gives rise to this feature.

\end{abstract}

\section{Introduction}
While the inner discs of most disc galaxies are remarkably flat, their outskirts are strongly warped. 
Consistent with expectations from simple approximations for vertical bending waves (Hunter \& Toomre 1969), the amplitude of these warps increases strongly with radius, reaching a level of few inner disk scale height.
The warps are conspicuous in the neutral gas plane as revealed by 21-cm H{\sc i} observations (e.g. Sancisi 1976; Bosma 1981; Briggs 1990) and also seen in the stellar plane through optical and infrared observations (e.g. Sanchez-Saavedra, Battaner \& Florido1990; Reshetnikov \& Combes 1998; Saha, de Jong \& Holwerda 2009).
As a typical spiral galaxy, the Milky Way also displays significant warps in the outer disk as revealed by various tracers, including neutral gas (e.g. Burke 1957; Kerr 1957), molecular clouds (e.g. Grabelsky et al. 1987), stars (e.g. Drimmel \& Spergel 2001; L{\'o}pez-Corredoira et al. 2002) and even interstellar dust (e.g. Freudenreich et al. 1994; Drimmel \& Spergel 2001).

Theoretically, the warp in a spiral galaxy are generally considered as a response of the galactic disk to a perturbation, which could come from the dark matter halo (assuming misaligned angular momenta; e.g. Ostriker \& Binney 1989; Debattista \& Sellwood 1999), the nearby satellite galaxies (such as the Sagittarius dwarf or the Large/Small Magellanic Clouds;  e.g. Weinberg 1995; Garcia-Ruiz, Kuijken \& Dubinski 2002; Bailin 2003), the cosmic infall (Jiang \& Binney 1999) or the intergalactic accretion flows (Lopez-Corredoira, Betancort-Rijo \& Beckman 2002; S{\'a}nchez-Salcedo 2006). 
Currently, the exact origin of the Galactic warp is still under hot debate, although so many mechanisms have been proposed for its generation.
The biggest challenge for solving this problem is to explore the dynamic nature of the Galactic warp on the basis of its well known structure.

At present, although several  attempts (e.g. Miyamoto, Yoshizawa \& Suzuki 1988; Dehnen 1998; L{\'o}pez-Corredoira et al. 2014; Poggio et al. 2017) have been made, { particular sub-groups or types of stars (i.e. OB stars or red clump stars)}, to study the kinematic behaviours of the Galactic warp, the results { no definite conclusions  have been reached} due to the  lack of sufficient proper motion data, either poor in accuracy for the ground-based or limited in number for space-borne {\it Hipparcos} (ESA 1997).
Recently, the first {\it Gaia} data release ({\it Gaia} DR1; Gaia Collaboration et al. 2016a) from {\it Gaia} survey (Gaia Collaboration et al. 2016b) has released accurate proper motion and parallax measurements of about two million stars by combining the data of Tycho and $Gaia$ (TGAS; Lindegren et al. 2016).
With this data set, Sch{\"o}nrich \& Dehnen et al. (2017; hereafter SD17) have estimated the vertical velocities $V_{z}$, azimuthal velocities $V_{\phi}$  and vertical angular momenta $L_{z}$ for stars in the Galactic center and anti-center directions.
They find that the mean vertical velocity $\overline{V_{z}}$ shows a clear increasing trend with both $V_{\phi}$ and {$L_{z}$}, as expected from a long-lived Galactic warp (e.g. Drimmel, Smart \& Lattanzi 2000; hereafter DSL00).
Moreover, a previously unknown wave-like pattern of $\overline{V_{z}}$ versus $L_{z}$ (and versus $V_{\phi}$ and guiding center radius $R_{g}$) with an amplitude of $\sim 1$\,km\,s$^{-1}$ on a scale of $\sim 2.5$\,kpc is detected.
They suggest that the feature could arise from a winding warp or bending waves.
In addition, a stream-like feature near $R_{g} \sim 9$\,kpc ({$L_{z} \sim 2150$\,kpc\,km\,s$^{-1}$}) is found to deviate significantly from this wave-like pattern.

\begin{figure*}
\begin{center}
\includegraphics[scale=0.435,angle=0]{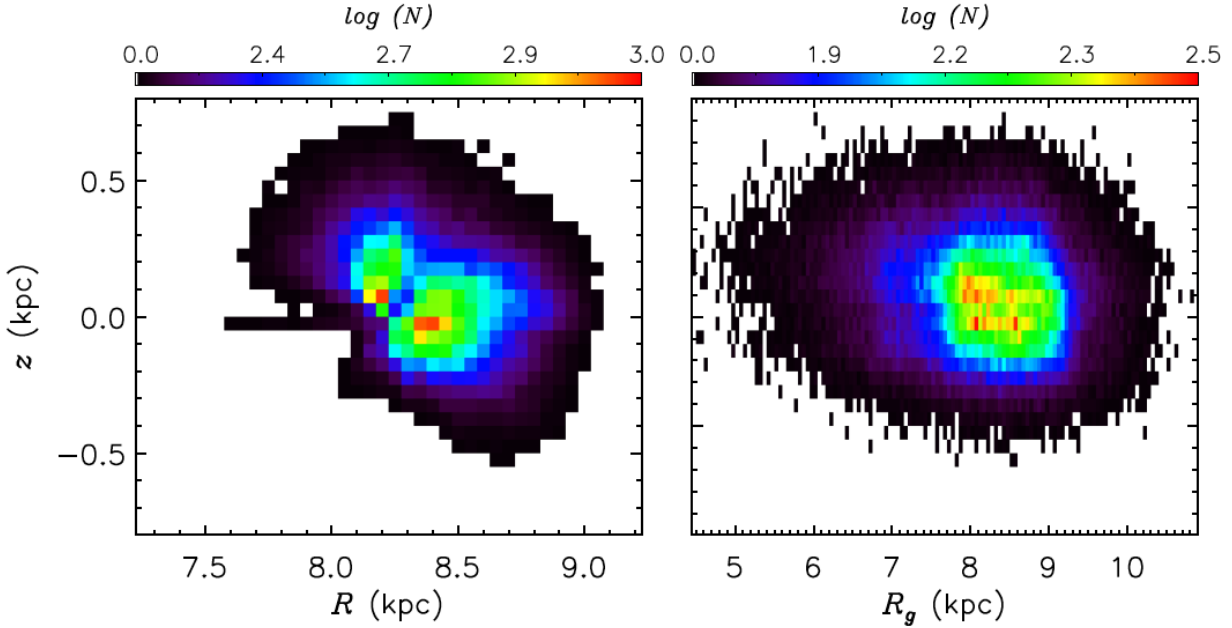}
\caption{Spatial distribution of the final LAMOST-TGAS common stars in the $R$--$z$ (left) and  $R_{g}$--$z$ (right) planes.
The stellar number densities (with binsize of 0.05\,kpc in each axis) are indicated by the top colorbars.}
\end{center}
\end{figure*}

\begin{figure}
\begin{center}
\includegraphics[scale=0.35,angle=0]{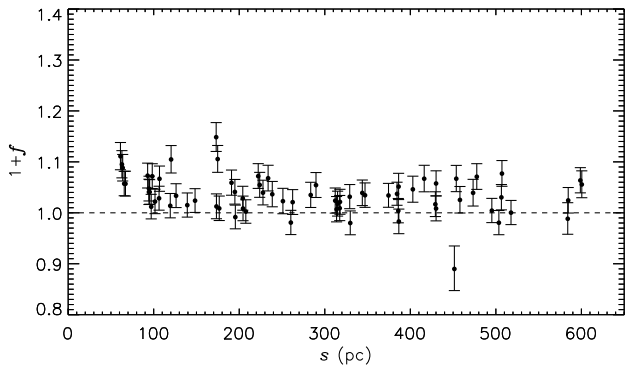}
\caption{ Distance estimator $1+f$ versus distance $s$ for the  the LAMOST-TGAS stars used in this work。
The stars are binned in distance such that each bin contains of 3000 stars , and the distance mask slides in steps of 1000 stars.
For the definition of estimator $1+f$, please refer to SBA. 
In principle, $1+f$ should be\,unity if there are no systematic biases in the estimated distances.}
\end{center}
\end{figure}

\begin{figure}
\begin{center}
\includegraphics[scale=0.55,angle=0]{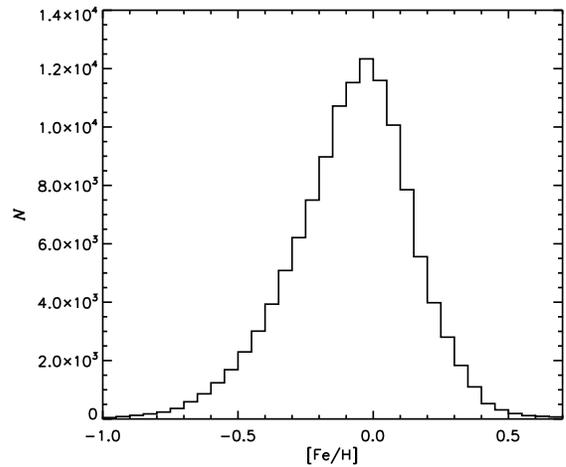}
\caption{ Metallicity distribution of the final selected 123\,233 LAMOST-TGAS common stars.}
\end{center}
\end{figure}

\begin{figure}
\begin{center}
\includegraphics[scale=0.45,angle=0]{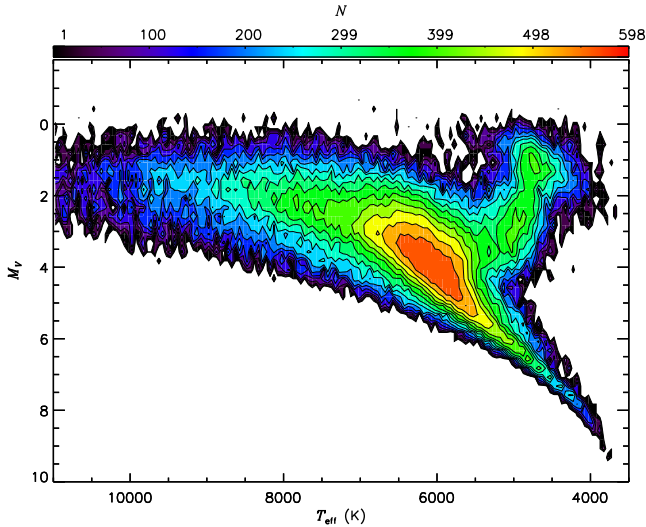}
\caption{ Pseudo-color $T_{\rm eff}$--$M_{\rm V}$ diagram of the final selected 123\,233 LAMOST-TGAS common stars.
                               The stellar number density is indicated by the top colorbar.}
\end{center}
\end{figure}

In this {paper}, we have repeated the analysis of SD17 using full three-dimensional (3D) velocity measurements (tangential plus line-of-sight velocities) by combining the LAMOST and TGAS data, rather than tangential velocities only as in SD17.
The accurate 3D velocity measurements for a large number of stars available from the LAMOST-TGAS data provide a direct check of the results of SD17.
The letter is organized as follows.
In Section 2, we briefly describe the LAMOST-TGAS data set. 
The results are presented and discussed in Section 3. 
Finally, we summarize in Section 4.

\begin{figure*}
\begin{center}
\includegraphics[scale=0.6,angle=0]{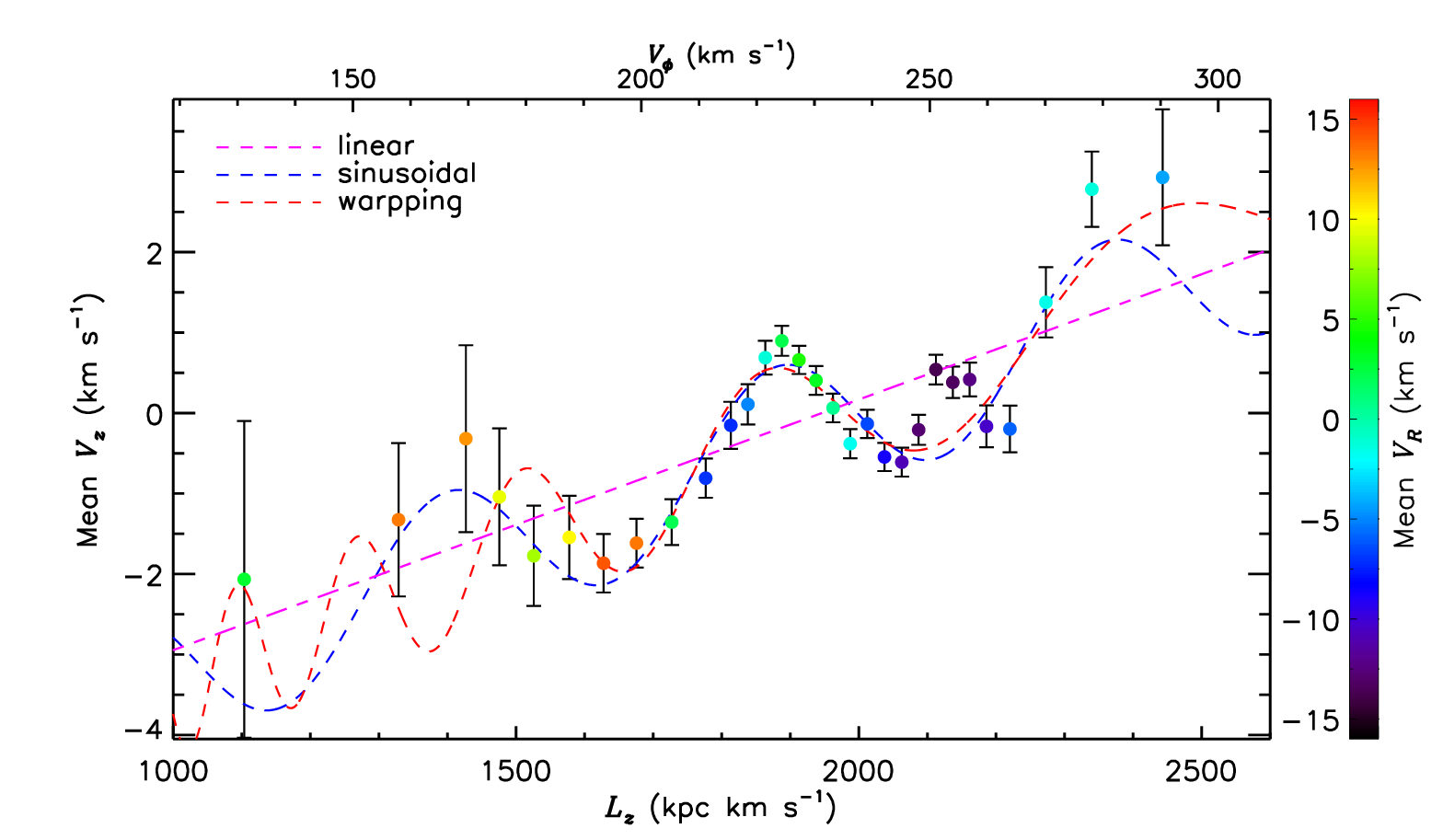}
\includegraphics[scale=0.6,angle=0]{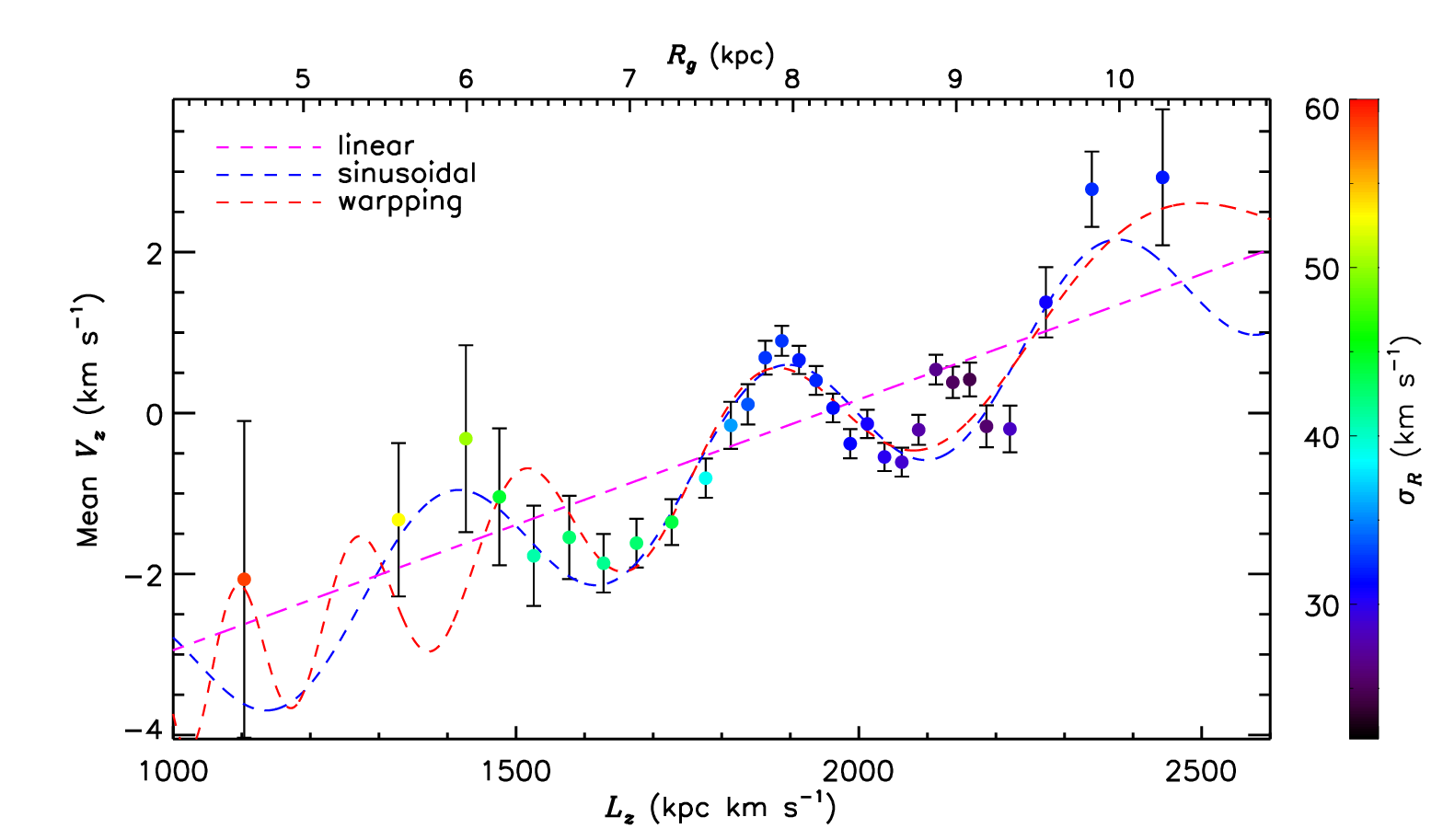}
\caption{Mean $V_{Z}$ velocity versus $L_{z}$ or azimutal velocity $V_{\phi}$ or guiding canter radius $R_{\rm g}$ for final LAMOST-TGAS sample.
              The  magenta, blue and red dash lines represent a linear fit from Equation\,(1), a simple sinusoidal  fit from Equation\,(2) and a wrapping fit from Equation\,(3), respectively, and the best-fit parameters are presented in Table\,1.
              The colors of the data points show the values of mean radial velocity $V_{R}$ (top panel) and radial velocity dispersion $\sigma_{R}$ (bottom panel) of each angular momentum bin, respectively.
              The binsize is set to be no smaller than $50$\,kpc\,km\,s$^{-1}$ but allowed to be larger to ensure no less than 100 stars in each bin.}
\end{center}
\end{figure*}

\begin{figure*}
\begin{center}
\includegraphics[scale=0.6,angle=0]{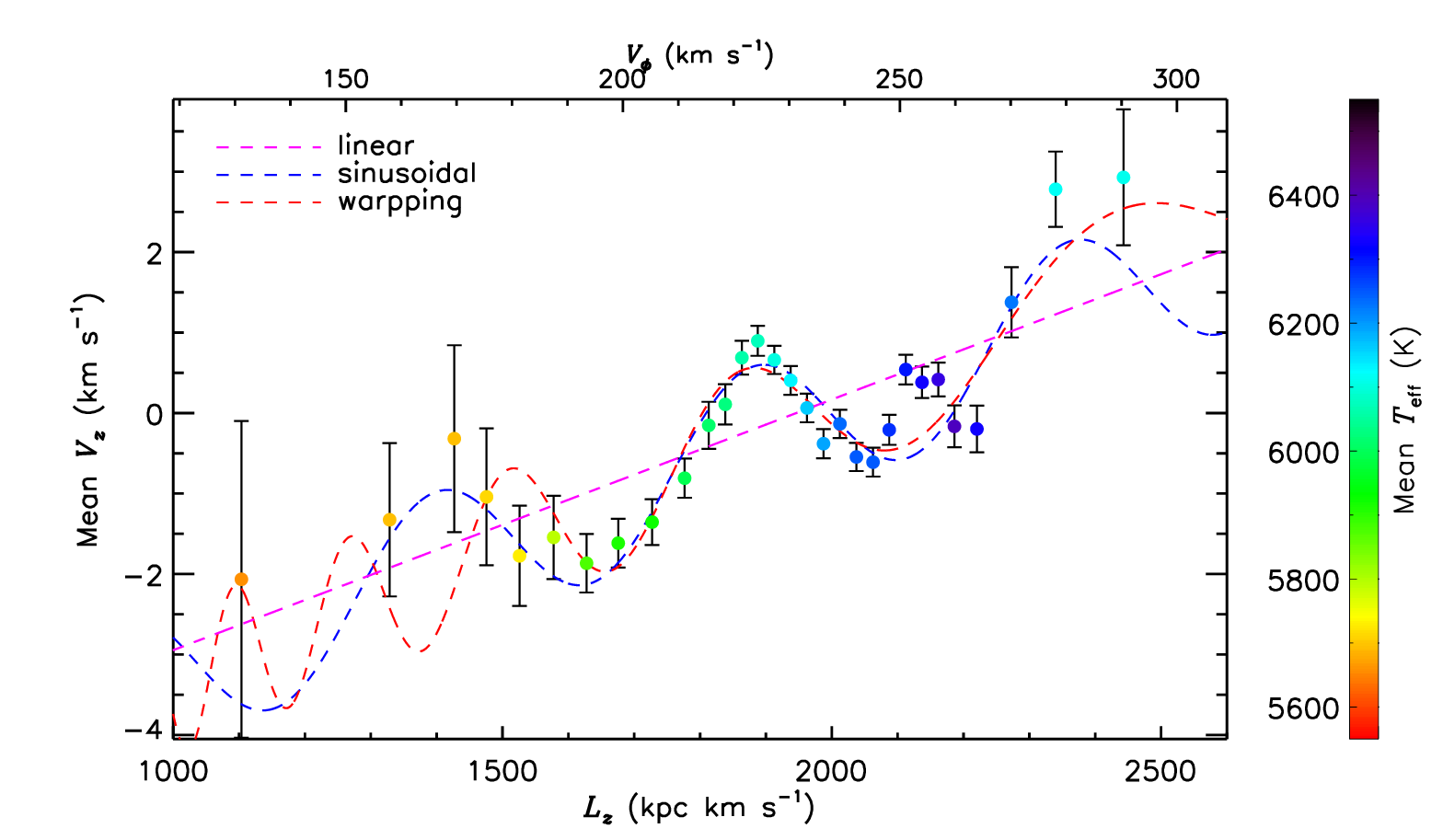}
\includegraphics[scale=0.6,angle=0]{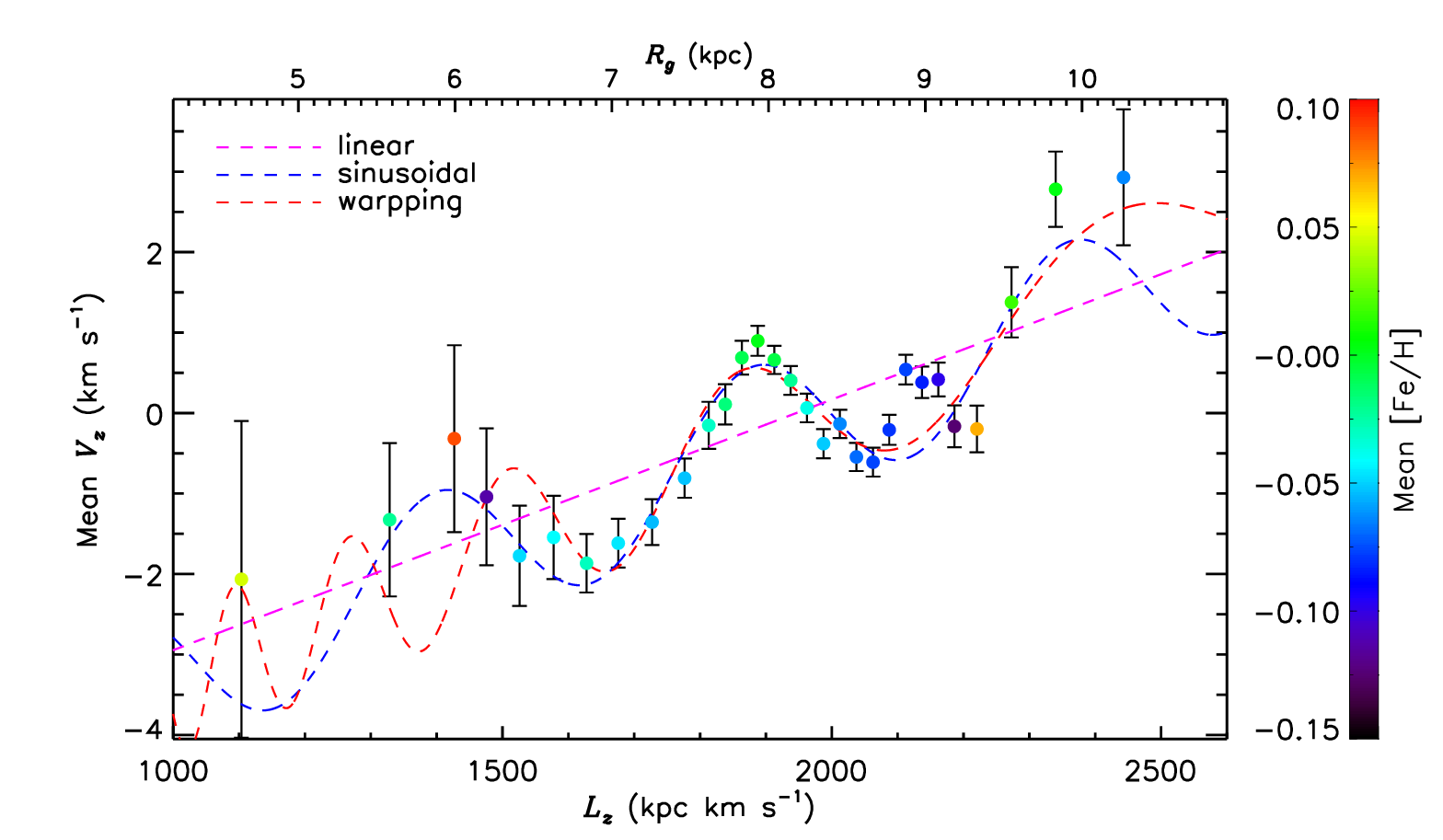}
\caption{Similar to Fig.\,5 but with colors coded by mean $T_{\rm eff}$ (top panel) and by mean [Fe/H] (bottom panel).}
\end{center}
\end{figure*}

\section{Data}

\subsection{Coordinate systems}
In this study, we use the standard Galactocentric cylindrical system ($R$,\,$\phi$,\,$z$) with $R$, the projected Galactocentric distance, increasing radially outwards;
$\phi$ in the direction of Galactic rotation and $z$ towards the North Galactic Pole.
The three velocity components are respectively represented by $V_{R}$,\,$V_{\phi}$,\,$V_{z}$.
The guiding center is defined as $R_{g} = RV_{\phi}/V_{c} (R_{0})$, where $R_0$ is the Galactocentric distance of the Sun, given the fact that the Galactic rotation curve in the solar neighborhood is nearly flat (e.g. {McMillan 2011}; Reid et al. 2014; Huang et al. 2016).
For the solar motions and $R_{0}$, we use the same values adopted by SD17, i.e. $V_{\phi, \odot} = 250$\,km\,s$^{-1}$, ($U_{\odot}$,\,$V_{\odot}$,\,$W_{\odot}$)\,$=$\,$(13.00,\,12.24,\,7.24)$\,km\,s$^{-1}$ and $R_{0}\,=\,8.27$\,kpc, determined by Sch{\"o}nrich, Binney \& Dehnen (2010) and Sch{\"o}nrich (2012), and consistent with results of Reid \& Brunthaler (2004), Reid et al. (2014) and Huang et al. (2015, 2016), resulting in  $V_{c} (R_{0})$ is $238$\,km\,s$^{-1}$.

\subsection{LAMOST and TGAS data}
In this work, we use the data from two surveys: the LAMOST Galactic spectroscopic surveys and the {\it Gaia} survey.

Full descriptions (including scientific motivations and target selections) of the LAMOST Galactic spectroscopic surveys are presented in Zhao et al. (2012), Deng et al. (2012) and Liu et al. (2014).
The LAMOST Phase-I Surveys were initiated in the fall of 2012 (after one year Pilot Surveys from September 2011 to June 2012) and completed in the summer of 2017.
The Phase-II  Pilot Surveys kicked off last September, with the Phase-II Regular ones expected to begin in the coming September for another period of five years.
At present, with the stellar parameter determination pipeline LSP3 (Xiang et al. 2015), line-of-sight velocities and basic atmospheric parameters (effective temperature $T_{\rm eff}$, surface gravity log\,$g$, metallicity [Fe/H]) have been derived from approximately 6.5 million qualified (i.e. of signal-to-noise ratios -- SNRs greater than 10) stellar spectra of about 4.4 million stars, collected with LAMOST by June 2016.
This data set will soon be publicly available as value-added catalogue for LSS-GAC DR3.

The European Space Agency (ESA) {\it Gaia} satellite was launched in December 2013.
It will obtain highly accurate parallaxes and proper motions for nearly one per cent ($\sim$\,one billion) of all stars in the Milky Way of a $G$ magnitude brighter than $\sim$\,$20.7$ (Gaia Collaboration et al. 2016b).
With early survey data, the {\it Gaia} DR1 (Gaia Collaboration et al. 2016a) was released to the community on September 14, 2016.
The astrometric TGAS catalogue provides parallaxes and proper motions for over two million stars from the Tycho and Hipparcos catalogues (Lindegren et al. 2016).

We have cross-matched the LSS-GAC DR3 with the {\it Gaia} TGAS and found nearly 0.25 million common stars of LAMOST spectral SNRs greater than 10.
To ensure accurate 3D velocities, we have further applied the following cuts to the LAMOST-TGAS sample of common stars:
\begin{itemize}[leftmargin=*]
\item  {LAMOST spectral SNRs greater than 30;}

\item  {metallicity [Fe/H]\,$\ge -1.0$\,dex;} 

\item {relative parallax errors smaller than 20 per cent in TGAS.}
\end{itemize}
The first cut ensures that the precision of the LAMOST line-of-sight velocities is better than 5\,km\,s$^{-1}$ (Huang et al. 2018).
Note that  line-of-sight velocities yielded by LSP3 have been corrected for a systematic offset of 3.1\,km\,s$^{-1}$ (Xiang et al. 2015).
The second cut is used to exclude potential contaminations from the halo stars.
On top of the third cut, we have adopted the distance estimates of Sch{\"o}nrich \& Amuer (2017; hereafter SA17), who consider both the distance prior and the survey selection function prior in deriving the distances from the TGAS parallaxes.
{ For the validation of the derived distances, we use the method of Sch{\"o}nrich, Binney \& Asplund (2012; hereafter SBA) and find that the potential systematic uncertainties are smaller than a few per cent (overestimated by $\sim$\,3.3 per cent, see Fig.\,2).
 The distance bias is small and the measurement may be affected by the residual uncertainties in the determination of LAMOST line-of-sight uncertainties. 
 More importantly, the statistics are very stable versus distance, affirming the quality of the dataset. 
 Our data are less prone to distance bias than a mere proper motion sample as used in SD17, and a 3 per cent distance offset should result in a general vertical velocity bias of less than 0.4 km/s, as also tested in our analysis.
 Therefore, we can refrain from further corrections.}
With the third cut and the SA17 technique, precise and bias free distances have been determined for 123\,233 LAMOST-TGAS common stars after applying the first two cuts.
With the above estimated distances, together with the line-of-sight velocities\footnote{{The LAMOST line-of-sight velocities are also tested bias free on the the level of 2\,km\,s$^{-1}$ by the SBA method.}} from the LSS-GAC DR3 and the proper motions from the TGAS, we further derive the 3D ($R, \phi, z$) positions, and 3D ($V_{R}$,\,$V_{\phi}$,\,$V_{z}$) velocities, for these stars.
Finally, the values of guiding center radius $R_{g}$ and vertical angular momentum $L_{z}$ are respectively calculated using $R_{g} =  RV_{\phi}/V_{c} (R_{0})$ and {$L_{z} = RV_{\phi}$}.

\begin{table*}
\centering
\caption{Best-fit parameters of three models, described by Equations (1-3), fit to the LAMOST-TGAS data as shown in Fig.\,2.}
\begin{tabular}{lccccc}
\hline
Fit&$a$ & $b$ & $c$ & $d$ & $A$ \\
&($\times 10^{3}$ kpc)&(km\,s$^{-1}$)&(kpc km\,s$^{-1}$)&&(km\,s$^{-1}$)\\
\hline
linear&$3.11 \pm 0.70$&$-1.078 \pm 0.273$&--&--&--\\
sinusoidal& $3.24 \pm 0.38$&$-1.277 \pm 0.149$&$481 \pm 20$&$-2.05 \pm 0.22$ & $0.95 \pm 0.12$\\
wrapping&$3.47 \pm 0.44$&$-1.283 \pm 0.174$ & $7951 \pm 341$& $-0.12 \pm 1.13$ & $0.90 \pm 0.14$\\
\hline
\end{tabular}
\end{table*}

The spatial distribution of the final LAMOST-TGAS sample of common stars is presented in Fig.\,1.
As the plot shows, the sample is restricted to within $\sim 0.6$\,kpc of the Sun, but guiding centre radii are probed in the range $5 < R_g < 10$\,kpc. 
The short distance range of the sample also restricts the vertical extent and thus limits contamination of our signal by breathing modes.
{The metallicity distribution of the final selected sample is shown in Fig.\,3.
The median of the distribution is around $-0.06$\,dex, indicating a thin disc population nature of  the sample. 
A plot of the sample in $T_{\rm eff}$--$M_{V}$ plane is shown in Fig.\,4.
Here the effective temperatures are estimated from color $(V-K_{\rm s})_{0}$, using the empirical metallicity-dependent calibration relations provided by Huang et al. (2015).
We have also adopted the values of the interstellar reddening of the individual stars derived with the ‘standard pair’ technique (Yuan, Liu \& Xiang 2013).}

\section{Results and discussion}
Systematic motions perpendicular to the Galactic plane will be induced if there is a long-lived warp in the stellar disk.
Typically, the systematic vertical motions show an increase trend with $R$ on large-scale (e.g. DSL00).
As mentioned above, our sample covers a wide range of $R_{g}$ but not $R$.
Following SD17, we therefore study the mean vertical motions of our sample as a function of $L_{z}$, and of $R_{g}$ and of $V_{\phi}$.

\subsection{The kinematic signature of the Galactic warp}
As in SD17, we examine the mean values of vertical velocity ${\overline V_{z}}$ of our final LAMOST-TGAS sample in the individual bins of vertical angular momentum $L_{z}$. 
The binsize in $L_{z}$  is allowed to vary to ensure no less than 100 stars in each bin but set to be no smaller than $50$\,kpc\,km\,s$^{-1}$.
The latter is to set to match with the typical uncertainties of $L_{z}$.
The error bar of ${\overline V_{z}}$ for a given bin is given by $\sigma_{V_{z}}/\sqrt{N}$, where  $\sigma_{V_{z}}$ is the measured vertical velocity dispersion and $N$ the total number of stars in that bin.
The results are shown in Fig.\,5, with $V_{\phi}$ (assuming a mean radius ${\overline R}$ of $\sim 8.41$\,kpc of the sample) and $R_{g}$ (assuming a flat rotation curve with a local circular speed of $238$\,km\,s$^{-1}$) also labeled. 

The overall mean vertical velocity ${\overline V_{z}}$ shows a significant increase and a wave-like pattern over {$ L_{z}$}, $V_{\phi}$ and $R_{g}$, in excellent agreement with the result derived by SD17 using the TGAS data only and also the expectation of a large-scale warp signal in the stellar disk (e.g. DSL00).
The trend in ${\overline V_{z}}$ versus $R_{g}$ shows a large-scale warp signal for $R_{g}$ ranging from about $6.0$ to 10.5\,kpc.
This places a strong constraint of the start position of the Galactic warp.
The onset radius of the galactic warp is highly debated, with estimates ranging from within, at or outside the solar annulus (Derriere \& Robin 2001; Drimmel \& Spergel 2001).
Our results place a strong constraint for the onset point, significantly within the solar annulus, thus in consistent with what found by Drimmel \& Spergel (2001) with the near-infrared photometry.

The feature at {$L_{z} \sim 2150$\,kpc\,km\,s$^{-1}$} (corresponding to $R_{g} \sim 9$\,kpc and $V_{\phi} \sim 255$\,km\,s$^{-1}$) is suspiciously narrow and likely a stream, so we censor the region from the following fits. 
We will return to this question in Section 3.3.

\subsection{Fits with simple models}

In this subsection, we attempt to fit our new data with three simple models first proposed by SD17.

If one assumes a simple, perfectly static, non-wrapped warp model (e.g. DSL00, Poggio et al. 2017), the correlation between  $\overline{V_{z}}$ and $L_{z}$  can be locally approximated by a linear fit:
\begin{equation}
\overline{V_{z}} = b + aL_{z}^{'} \text{,} 
\end{equation}
where { $L_{z}^{'} = L_{z} - 1600$\,kpc\,km\,s$^{-1}$}.

The above pure linear relation only described the general trend of increase of $\bar{V_z}$ with $-L_z$ but not the wave-like pattern.
For the latter, two simple yet physically motivated models are considered: 1) The disk may oscillate vertically as radially propagating waves (namely bending waves; Hunter \& Toomre 1969); and 2) The Galactic warp is excited by perturbations of dwarf satellites (e.g. the Sagittarius or the Magellanic Clouds) falling into the Galactic halo (namely a wrapping/winding warp; Weinberg 1995; Weinberg \& Blitz 2006). 
For the former, one has the following simple relation,
\begin{equation}
\overline{V_{z}} = b + aL_{z}^{'}  + A\sin(2\pi L_{z}^{'}/c + d) \text{,} 
\end{equation}
where $A$ is the amplitude of the wave-like pattern, $c$  the period and $d$ the phase.
For the second model, the high stellar vertical oscillation frequencies compared to the Magellanic Clouds orbital frequencies lead the warp wrapping up, with the warp signal decreasing toward larger radii.
Naively, we have the following relation for this model,
\begin{equation}
{\overline{V_{z}} = b + aL_{z}^{'}  + A\sin(2\pi c/L_{z} + d) \text{.} }
\end{equation}

We fit the measured $\overline{V_{z}}$--$L_{z}$ variation with the three simple models described above.
The fitting results are shown in Fig.\,5, with the best-fit parameters presented in Table\,1.
As the plot shows, the linear model describes the general trend of increase very well, but not the wave-like pattern.
The other two models describe both perfectly.
The amplitude of the wave-like pattern yielded by the latter two models is about $0.9$\,km\,s$^{-1}$, similar to what found by SD17.
The radial wavelength of oscillation yielded by the simple sinusoidal model is about $2.0$\,kpc, again in excellent agreement with the result of SD17.

We note that one should not over-interpret the later two naive fits considering that the spatial coverage of the current sample is too limited to examine the underlying physics of both models.
In addition,  the range is too small to detect any changes in radial wave-length with radius, which is exemplified by both bracketing models working quite well.
The future {\it Gaia} release, together with spectroscopic data (e.g. from LAMOST), will help solve this problem.

\subsection{A new stellar stream?}
Finally, we discuss the prominent localized peak at {$L_{z} \sim 2150$\,kpc\,km\,s$^{-1}$} found here and earlier by SD17.
As shown in Fig.\,5, the peak deviates significantly from all fits.
To understand the nature of this peak, we show the mean radial velocity $\overline{V_{R}}$ and radial velocity dispersion $\sigma_{R}$ in the individual $L_{z}$ bins in Fig.\,2.
As the plot shows, stars near the peak region have a significant { inward} mean radial motion (i.e. toward Galactic center direction) of $-13.33 \pm 0.59$\,km\,s$^{-1}$, while those outside the peak have motions are close to zero.
Stars near the peak region have a small radial velocity dispersion of only $25.27 \pm 0.89$\,km\,s$^{-1}$, significantly smaller than the local radial velocity dispersion ($\sigma_{R0} \sim 35$\,km\,s$^{-1}$; e.g. Huang et al. 2016).
In addition, as Fig\,6 shows, stars near this region also show a significant $T_{\rm eff}$ peak ($\sim 6300$\,K, about 100\,K higher than the nearby bins) and have a [Fe/H] dip ($\sim - 0.08$, 0.05-0.10\,dex lower than the nearby bins).
These results strongly suggest that this localized peak arises from a young and cold stellar stream.

\section{Summary}
Using over a hundred thousand stars with accurate 3D velocity measurements selected from the LAMOST-TGAS data, we have repeated the work of SD17 (using tangential velocities from the TGAS only) to detect kinematic signature of the Galactic warp.

With this data, an general trend of increase trend combined with a wave-like pattern of mean vertical velocity $\overline{V_{z}}$ versus absolute vertical angular momentum $L_{z}$ and azimuthal velocity $V_{\phi}$ is clearly detected for guiding center radius $R_{g}$ between $\sim$\,6.0 and 10.5 kpc.
The increase is expected from a {large-scale} Galactic warp.
The wave-like pattern with an amplitude of $\sim 0.9$\,km\,s$^{-1}$ on a scale of $\sim 2.0$\,kpc may arise from bending waves or a winding/wrapping warp.

In addition to the above trend and wave-like pattern, a prominent localized peak is detected near {$L_{z} \sim 2150$\,kpc\,km\,s$^{-1}$} (corresponding to $R_{g} \sim 9$\,kpc  and $V_{\phi} \sim 255$\,km\,s$^{-1}$).
It has a quite large { inward} radial velocity of $-13.33 \pm 0.59$\,km\,s$^{-1}$ and a small radial velocity dispersion of $25.27 \pm 0.89$\,km\,s$^{-1}$, suggesting a cold stellar stream nature of this feature.

In summary, this work and SD17 confirm an increase trend associated with a wave-like pattern in $V_{z}$ over {$L_{z}$}.
The result should provide vital constraints to future disc warp and (vertical) oscillation modeling.

 \section*{Acknowledgements} 
The Guoshoujing Telescope (the Large Sky Area Multi-Object Fiber Spectroscopic Telescope, LAMOST) is a National Major Scientific Project built by the Chinese Academy of Sciences. Funding for the project has been provided by the National Development and Reform Commission. LAMOST is operated and managed by the National Astronomical Observatories, Chinese Academy of Sciences.
The LAMOST FELLOWSHIP is supported by Special fund for Advanced Users, budgeted and administrated by Center for Astronomical Mega-Science, Chinese Academy of Sciences (CAMS).
R.S. is supported by a Royal Society University Research Fellowship.

{ This work has made use of data from the European Space Agency (ESA) mission Gaia (https://www.cosmos.esa.int/gaia), processed by the Gaia Data Processing and Analysis Consortium (DPAC, https://www.cosmos.esa.int/web/gaia/dpac/consortium).}

{This work is supported by the National Key Basic Research Program of China 2014CB845700, the China Postdoctoral Science Foundation 2016M600849 and the National Natural Science Foundation of China U1531244 and 11473001.  
It is pleasure to thank Zuhui Fan for valuable discussions.}

\end{document}